\documentstyle[editedvolume,psfig]{crckapb} 

\begin{opening}
\title{RELATIVISTIC MODELS OF kHz QPOs}

\author{W. KLU\'ZNIAK}
\institute{Copernicus Astronomical Center\\
           ul. Bartycka 18, 00-716 Warszawa, Poland\\
           wlodek@camk.edu.pl}

\end{opening}

\runningtitle{General Relativity and quasiperiodic oscillations}

\begin{document}


\begin{abstract}
After reviewing the general-relativistic ``gap'' model of accretion,
I discuss its relation to the high frequency quasi-periodic oscillations
observed in low-mass X-ray binaries. The ``300'' Hz frequency seen
in some X-ray bursts may be a relativistic signature of keplerian rotation
of the neutron star.
\end{abstract}

It is easy to see how much the field has advanced in the past decade
by comparing the topics under discussion here in Elounda, with those
discussed at that previous conference of this series which
also took place in Crete, in Agia Pelagia, at the beginning of the decade. 

Back then,
magnetic fields were all the rage. Gamma-ray bursts supposedly
showed in their spectra cyclotron absorption lines suggesting
(to many) that the sources are Galactic neutron stars, a view
completely ruled out in the decade of Compton GRO, Beppo SAX
and the observations of afterglows (described carefully in Dr. Fishman's
talk here in Elounda). Another view much discussed at Agia Pelagia
was that low-mass X-ray binaries contain direct counterparts of
millisecond pulsars, i.e., $10^9$ to $10^{10}\,$Gauss neutron stars,
rotating at periods of a few milliseconds, and therefore
accreting through a disk in which the orbital frequency  (supposedly) differed
from the stellar rotational frequency by about 50 Hz. Today, after the
discovery of kHz QPOs (described in these proceedings by M. van der Klis),
there is hardly any doubt that the characteristic orbital
frequency in the inner
disk is at least 1 kHz, very different from the value of about
300 Hz promoted at Agia Pelagia. 

So let us forget about magnetic fields in LMXBs and ask what would be
expected then. The answer depends on the equation of state (e.o.s.) of matter
at supranuclear densities and on the mass of the neutron star, as
well as its angular momentum. Here I will only discuss the general
relativistic ``gap'' regime of accretion, in which the accretion
disk does not extend to the stellar surface---there are good reasons for
that.

\section{The relativistic gap regime}.

 LMXBs are old accreting systems, so {\it a priori} one would
expect that the central neutron stars have each gained a few tenths of a solar
mass since their early days. Now, as pointed out sime time ago
(Klu\'zniak and Wagoner 1985), for all e.o.s., at sufficiently high
stellar mass (which need not be very large), a slowly rotating
neutron star is within
the innermost stable circular orbit (ISCO) allowed by general relativity (GR),
 a.k.a.  the marginally stable orbit.
For rapidly rotating neutron stars this is not always so, but
according to the tables of Cook et al. (1994), for
most e.o.s. the maximally rotating models are also within the ISCO.
For strange (quark) stars
this is also true (Stergioulas et al. 1999).
In short, it seems reasonable to assume that in LMXBs, the compact object
is inside the ISCO, so let us do so.

The three-dimensional flow in accretion disks is still poorly understood
(it may resemble the flow of waves crashing on the beach, particularly
the rip tide dreaded by ocean swimmers---see the figure from Kita's 1995 thesis
reproduced in Klu\'zniak 1998b). But in any case,
in the relativistic gap regime, the disk should be terminated by GR effects,
as in the black hole  disks, whose essential properties were discussed
in numerous papers, e.g., of the Warsaw school some two decades ago
(by Paczy\'nski, Abramowicz, Sikora, Muchotrzeb and others, in various
combinations). Without further ado, let us accept the view that the
maximum observable frequency in LMXB disks is close to the ISCO frequency
and that this frequency may modulate the X-ray flux
(Klu\'zniak et al. 1990). Then it will be easy to believe
that the saturation (at 1.07 kHz) of QPO frequency in 4U 1820-30
is a signature of the
ISCO (Zhang 1998), and that the e.o.s. is severely constrained
by the observed maximum frequency value (Klu\'zniak~1998a).

What happens to the matter which leaves the disk through its inner edge
(assumed to be close to the ISCO radius)? It goes into free-fall and
approaches the surface at a rather shallow angle.
Under these conditions, a sheared atmosphere heated by the incoming
fluid is set up in the equatorial regions
 (or even the tropics, as in Dr. Sunyaev's talk), whose
vertical structure 
has been found in a 1+1--d calculation with full radiative transfer
(Klu\'zniak and Wilson 1991): the atmosphere is hot
and gives off radiation with a power law spectrum extending to about 200 keV.
This would agree with reports of hard radiation from several X-ray bursters. 
Of course this spectrum may be downgraded as the radiation interacts with
the (relatively) cool disk and the accretion stream, this interaction
has not yet been computed, but it is clear that on such a picture one
would expect the down-scattered softer photons to lag in time the harder
photons (as has been reported in SAX J 1808.4--3658).

My prejudiced view is that the balance of observations is in favour
of the gap regime. It really seems that several separate facts
(especially these: the presence of kHz QPOs and their frequency values,
 hard spectra, soft lags) suggest that the disk is terminated outside
the stellar surface by effects of general relativity.

\section{Model independent conclusions about QPOs?}

The power spectrum of both neutron star and candidate black hole systems
has been studied over the whole range of observed frequencies,
and the phenomenology of QPOs in both types of systems was found
to be remarkably similar (Psaltis et al. 1999). The neutron star
systems show two kHz QPOs, the one with lower frequency has
a clear counterpart in black hole sources (for example, in both types
of systems it has identical correlations with lower
 frequency features in the spectrum), only the highest frequency QPO 
 has a different phenomenology (Psaltis et al. 1999).

If the black hole candidates are indeed black holes, all black hole QPOs
and their counterparts in neutron star systems
must be accretion disk phenomena, reflecting fundamental properties
of flow in the  gravitational field of the compact object.
A logical conclusion would be that in the neutron star systems
it is the lower frequency kHz QPO which could
be connected with orbital motion, with its characteristic cut-off
frequency in the ISCO. This would make constraints on the e.o.s.
even more stringent than the ones inferred from the higher frequency
QPO, and discussed in the previous section
(but in either case, these constraints would be relaxed
if the QPO frequency were lower than the orbital frequency).
The same model should then describe the power spectra of neutron stars
in LMXBs as of accreting black holes, including those in AGNs (where
the frequency would be scaled down in inverse proportion
to the black hole mass).

On this picture, it would be the highest frequency kHz QPO alone,
which would need a special explanation for neutron star systems.
A special topic of attention must be the similar value
of the difference in the two ``kHz'' QPO frequncies, to the $\sim 300\,$Hz
frequency of the coherent peak
in the power spectrum seen in X-ray bursts (Strohmyer et al. 1997).

\section{Keplerian rotation?}

While accreting mass in LMXBs, the neutron stars are also accreting
angular momentum, a lot of it. Exact models (Cook et al 1994)
show that maximally rotating neutron stars have angular momentum
$J\approx 0.6 GM^2/c$, this amount of momentum can be accreted
already with $\sim 0.2M_\odot$ in mass (Klu\'zniak and Wagoner 1985).
Several instabilities are known which can limit the spin rate of a neutron
star (mostly through emission of gravitational radiation), but it is not known
whether they actually operate in practice. The most recently discovered
r-mode instability could, in principle, limit neutron star periods
to values even as long as a few milliseconds in LMXBs (Andersson et al. 2000).

On the observational side, there is no compelling evidence of the
periodicity of persistent accretors in LMXBs (with the exception of
the one or two strongly magnetized X-ray pulsars which have a low
mass companion, but have nothing to do with the atoll, banana, and other
LMXBs sources so colourfully described by Michiel van der Klis).
This is why the 2.5 ms coherent period discovered in the transient
 SAX J 1808.4--3658 gave rise to much excitement. Of course,
the famous ``300 Hz periodicity''  discovered in several
X-ray bursts has been interpreted as the stellar rotational
frequency (Strohmyer et al. 1997), but there seems to be
no good model for its observed properties
(the ``hot spot model'' has been criticized here in Elounda,
on different grounds, by Fred Lamb and Rashid Sunyaev).
The argument for rotation seems to be: what else could it be?
 Clearly there is a clock in
the system with a very good memory of frequency, and yet one which
wanders on short time scales.

The coherent peak in the power spectrum of bursters appears during the
 X-ray  burs, and  persists for several seconds, during
which the frequency increases usually, and yet from burst to burst
the frequency is amazingly stable. Usually these are hallmarks
of an (anharmonic)  oscillator. Can we find one in the system?

Marek Abramowicz and I think that we have found an anharmonic oscillator
on the road to Knossos, at least in our minds. In the equatorial
plane outside of (even a spinning) gravitating body the description of 
test-particle motion can be reduced
to one dimensional motion in an effective potential, $V$.
 The characteristic shape
of the effective potential in GR is shown in the figure. Stable
motion in circular orbits, as in the Newtonian case, is possible in 
the minimum of the potential---for the metric and angular momentum chosen
in the figure, this orbit would have radius $r_o=8GM/c^2$; for a different
angular momentum of the test particle in the same metric the potential
would have a minimum at a different radius, but always outside the ISCO,
which has a radius $r_{ms}$ uniquely fixed by the metric.
A characteristic feature of GR metrics is the existence of a maximum of
the effective potential (in the figure at $r_u=5GM/c^2$), at which unstable
circular motion is possible. In all cases, $r_u\le r_{ms}\le r_o$,
with the equality occuring at the minimum value of angular momentum
possible in circular motion of a test particle in the (fixed) metric.

\begin{figure}
\psfig{width=0.95\columnwidth,file=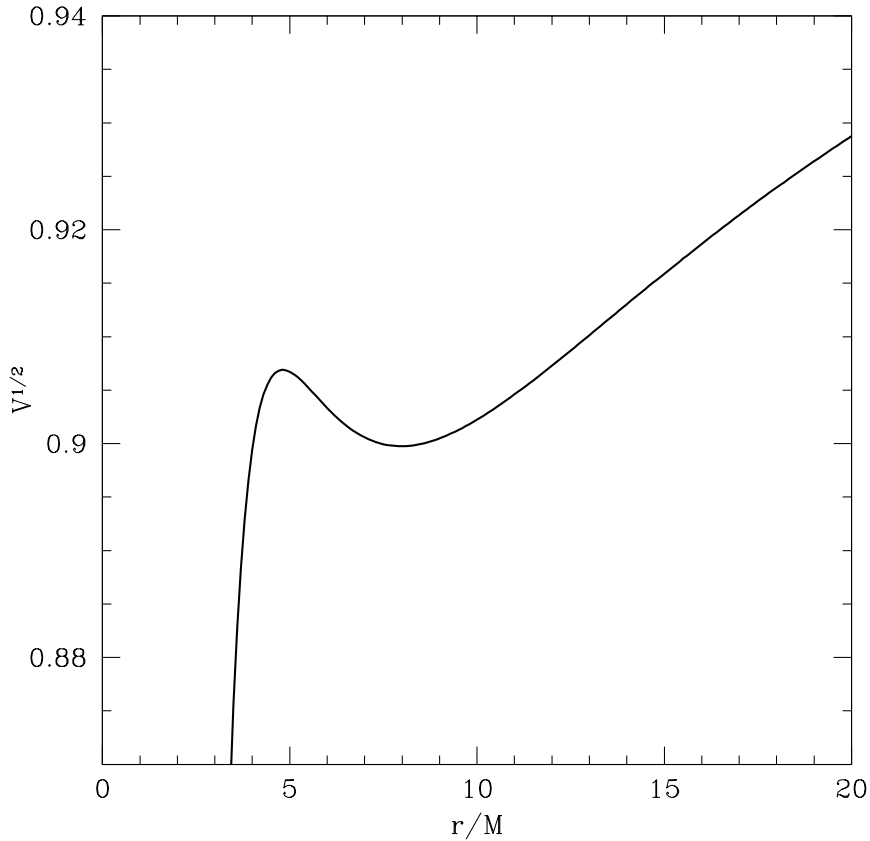}
\caption{The effective potential in GR (the detailed shape
of the curve will vary with the metric, here 
a Schwarzschild potential is shown).}
\end{figure}

The maximum (``Keplerian'') rotation rate of a star with radius $R$ inside
the ISCO (i.e., $R<r_{ms}$) occurs, when $R=r_u$. Imagine, then, 
that the effective potential in the figure
is that of test particles in the external
metric of a maximally rotating neutron star of radius $R=5M$,
or just a little bit less, for the same  value of
specific angular momentum as that of matter on the equator.
Now imagine that an explosion (an X-ray burst) lifts some matter
off the surface. The radial motion of the matter is an oscillation in the
potential well. If the energy of the matter were constant, it would
travel to a turning point (at about $r=12M$) and return to
the maximum of the potential. However, if a little energy is removed,
the matter oscillates back and forth between the right turning point,
and the left one, just below the maximum of $V$. This is the anharmonic
oscillator suggested as the origin of the ``$300\,$Hz burst oscillation,''
 with the frequency increasing as the amplitude of motion
(range in $r$) decreases. At last, the matter settles in circular orbit
at the bottom of the potential well.

\subsection{Prediction}

Clearly, the model presented  here requires maximal rotation of the neutron
star, which is expected to be higher than 600 Hz (because two radio pulsars
with 1.6 ms periods have aleady been observed).

We (Marek and I), would then predict that no X-ray burst oscillation
will be seen in X-ray bursts of sources with a clearly detected (phase
connected solution) rotational period of more than 1.6 ms.
For instance, in the transient SAX J 1808.4--3658, where a period of $P=2.5$ ms
has been measured, no such burst oscillation should be discovered.

\section{Summary}
It seems that the relativistic gap regime---the expected basic mode
of accretion onto neutron stars
with very weak magnetic fields---fits
most observations of LMXBs, including the essential phenomenology of
QPOs. The regime allows for neutron star rotation rates higher than
orbital frequencies in the disk. For maximally rotating neutron
stars, an oscillation in the relativistic ``potential well'' is
possible, with properties similar to those of the ``$300\,$
Hz'' oscillation observed in some X-ray bursts.
\smallskip~\\
This research was supported in part by KBN grant 2 P03D01816.

~\\Chryssa and Jan, thank you for the conference, we missed you,
 and we will always miss Jan.

\end{document}